\documentclass[journal]{IEEEtran}
\usepackage{graphicx}
\usepackage{amssymb}
\usepackage{multirow}
\usepackage{indentfirst}
\usepackage{inputenc}
\usepackage{array}
\usepackage{multicol}
\usepackage[switch,mathlines]{lineno}
\usepackage{soul}
\usepackage{graphicx}
\usepackage{cite}
\usepackage[cmex10]{amsmath}
\usepackage{algorithmic}
\usepackage{array}
\usepackage{CJK}
\usepackage[cmex10]{amsmath}
\usepackage{algorithmic}
\usepackage{array}
\usepackage{bm}
\usepackage{amssymb}
\usepackage{color}
\usepackage{nomencl}
\usepackage{pdfpages}
\usepackage{booktabs} 
\usepackage{multirow}
\usepackage{threeparttable}
\usepackage[ruled, linesnumbered]{algorithm2e}
\usepackage{pifont}
\usepackage{adjustbox}
\usepackage[colorlinks, linkcolor=blue, citecolor=blue, urlcolor=blue]{hyperref}
\usepackage[colorlinks, citecolor=blue]{hyperref}
\ifCLASSOPTIONcompsoc
 \usepackage[caption=false,font=normalsize,labelfont=sf,textfont=sf]{subfig}
\else
 \usepackage[caption=false,font=footnotesize]{subfig}
\fi
\usepackage{url}
\begin{document}

\title{A Survey of Open-Source Power System Dynamic Simulators with Grid-Forming Inverter \\for Machine Learning Applications}

\author{{Tong Su, \IEEEmembership{Student Member, IEEE}, Jiangkai Peng, \IEEEmembership{Member, IEEE}, Alaa Selim, \IEEEmembership{Student Member, IEEE}, \\ Junbo Zhao, \IEEEmembership{Senior Member, IEEE}, Jin Tan, \IEEEmembership{Senior Member, IEEE}}

\thanks{This work was authored by the National Renewable Energy Laboratory, operated by Alliance for Sustainable Energy, LLC, for the U.S. Department of Energy (DOE) under Contract No. DE-AC36-08GO28308. This material is based upon work supported by the U.S. Department of Energy's Office of Electricity under the OPEN COG Grid project.
\par Tong Su, Alaa Selim, and Junbo Zhao are with the Department of Electrical and Computer Engineering, University of Connecticut, Storrs, CT 06269, USA (e-mail: tongsu@uconn.edu; alaa.selim@uconn.edu; junbo@uconn.edu).
\par Jiangkai Peng and Jin Tan are with National Renewable Energy Laboratory, Golden, CO 80401, USA (e-mail: jiangkai.peng@nrel.gov; jin.tan@nrel.gov).
}}
\maketitle
\begin{abstract}
The emergence of grid-forming (GFM) inverter technology and the increasing role of machine learning in power systems highlight the need for evaluating the latest dynamic simulators.
Open-source simulators offer distinct advantages in this field, being both free and highly customizable, which makes them well-suited for scientific research and validation of the latest models and methods. This paper provides a comprehensive survey and comparison of the latest open-source simulators that support GFM, with a focus on their capabilities and performance in machine-learning applications.
\end{abstract}

\begin{IEEEkeywords}
    Open-source power system dynamic simulator, grid-forming inverter, machine learning application.
\end{IEEEkeywords}
\vspace{-0.3cm}
\section{Introduction}
\IEEEPARstart{T}{he} effective implementation of machine learning models, such as deep learning (DL) and reinforcement learning (RL) relies heavily on the availability and quality of the data. As machine learning methods gain prominence in power system studies, the role of data has become increasingly critical. Given the challenges of obtaining real-world power system data, most research relies on data generated by power system simulators. This is especially true for time-domain simulations (TDSs), as existing power grids operate far from the stability boundaries, making collecting sufficient data on instability events challenging. In addition, there is always a risk of perturbations or faults occurring at various locations for large power systems. Dynamic simulators can provide a detailed and interactive power system model, enabling engineers to anticipate and analyze various fault scenarios and system responses. This includes simulations ranging from simple single-device faults to complex interactions involving multiple devices. Through these simulations, the future safety state of the system and the effectiveness of control strategies can be assessed \cite{lara2023revisiting}. Machine learning methods require a large amount of data from power system simulators to accurately model power system dynamics and implement control following perturbations \cite{ibrahim2020machine, chen2022reinforcement}.

Power system simulators can be categorized into commercial and open-source simulators. Commercial simulators typically require a license but offer comprehensive user documentation, extensive functionality, highly specialized tools, as well as frequent software updates and feature upgrades. However, the advantage of open-source simulators is that they are usually free and highly customizable. Users can modify and improve the code according to specific needs, and new methods and technologies may be adopted more quickly. For example, ANDES is significantly easier to use for developing differential-algebraic equation (DAE)-based power system dynamic models compared to most commercial simulators, while maintaining high numerical efficiency \cite{cui2020hybrid}. Considering the ease of access and flexible scalability, open-source dynamic simulators are essential for academic research.

The existing power system simulator includes functionalities, such as power flow (PF), continuation power flow (CPF), optimal power flow (OPF), small-signal stability analysis (SSSA), and TDS. PF, CPF, and OPF are primarily based on static models to analyze the steady-state behavior under specific load and generation conditions. By contrast, SSSA and TDS utilize dynamic models to evaluate the system’s dynamic response and stability following perturbations. Static models-based simulators have been well-developed, like MATPOWER \cite{zimmerman2010matpower}, Pandapower \cite{thurner2018pandapower}, PowerSimulations.jl \cite{lara2021powersystems}, PyPSA \cite{PyPSA}, PowerModels \cite{coffrin2018powermodels}, PowerModelsDistribution.jl \cite{fobes2020powermodelsdistribution}, OpenDSS \cite{dugan2011open} and GridLAB-D \cite{chassin2014gridlab}. Among these, PowerModelsDistribution.jl, OpenDSS and GridLAB-D are primarily used for distribution systems. Due to the evolution of dynamic model structures and control methods, along with the level of modeling details required to capture specific phenomena of interest, there are variations in the support for dynamic models across different simulators. \cite{bam2005power, vogt2018survey, milano2005open} have conducted comprehensive surveys of various power system dynamic simulators, such as power system toolbox (PST) \cite{chow1992toolbox} and power system analysis toolbox (PSAT) \cite{milano2005open}. However, with the evolution of power systems and the emergence of new models, such as those for inverter-based resources (IBRs), there is a need to conduct surveys on the latest dynamic simulators in recent years.

This paper first introduces the definitions of TDSs, IBRs, and machine learning applications. It then surveys five leading open-source power system dynamic simulators that support grid-forming (GFM) inverters. Numerical studies are conducted on the IEEE 14-bus system to compare the performance of the simulators with different GFM controls.

\section{Power System Dynamic Simulation and Machine Learning Applications}

\subsection{Time-Domain Simulation}
\begin{figure}[htb]
  \centering
  \includegraphics[width=6.0cm]{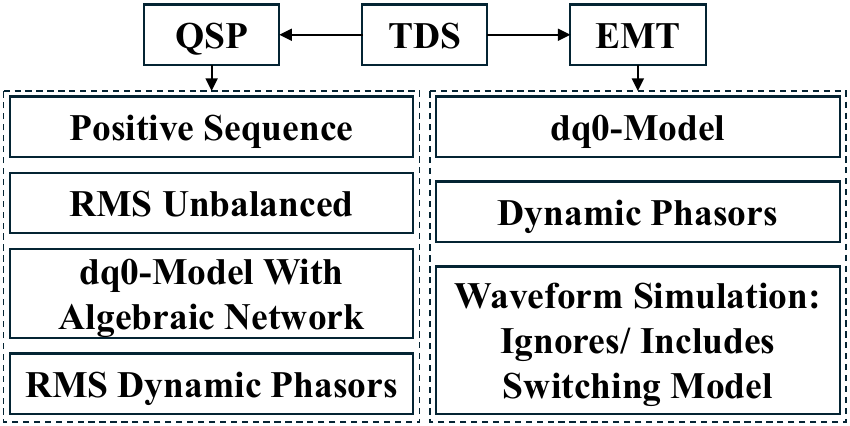}
  \caption{Classification of power system TDS.}
  \label{TDS}
\end{figure}

TDS is categorized into quasi-static phasor (QSP) simulation and electromagnetic transient (EMT) simulation, distinguished by the simulation speed and the level of modeling details necessary for capturing specific phenomena of interest. QSP simulations represent the dynamics of transmission system circuits as discrete changes between steady-state operating points and are used for studying low-frequency phenomena, ranging from inertial response to frequency regulation. EMT simulations incorporate sufficient detail to capture fast dynamic phenomena, enabling the analysis of line dynamics, converter switching, machine fluxes, and lightning surges \cite{lara2023revisiting}. Furthermore, QSP simulation is divided into the positive sequence, root mean square (RMS) unbalanced, dq0-model with algebraic network, and RMS dynamic phasors. On the other hand, EMT is categorized into dq0-model, dynamic Phasors, and waveform simulation (ignores or includes switching model). The classification of TDS is shown in Fig. \ref{TDS}.

\subsection{Inverter-Based Resource}
Traditional inverters use grid-following (GFL) control, relying on the phase-locked loop (PLL) to estimate and track the angle of the terminal voltage and regulate the injected current into the grid. However, GFL inverters, while unable to actively control voltage and frequency, behave similarly to a current source. When facing a contingency, GFL will act with a delay and cannot mimic the instantaneous response of synchronous generators (SGs), which leads to poorer system response performance \cite{tayyebi2020frequency}. On the other hand, GFM can directly control their voltage and frequency output to inherently enhance system stability and respond to load variations \cite{pattabiraman2018comparison}. The GFM source acts as a controllable voltage source and does not need the stiff terminal voltage assumption made in GFL \cite{matevosyan2019grid}. In emergencies like grid failure or islanded mode, GFM inverters establish the grid voltage, which then serves as the reference voltage for the rest of the system.

\begin{figure}[htb]
  \centering
  \includegraphics[width=7.6cm]{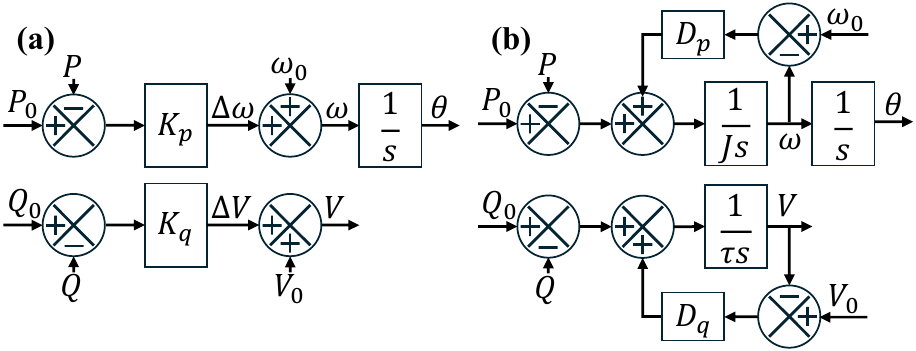}
  \caption{Structure of the GFM control: (a) Droop control. (b) VSM control.}
  \label{GFM}
\end{figure}

Several GFM controls have been proposed in recent years. Droop control, which mimics the speed droop mechanism present in SGs is widely used. Then, virtual synchronous machine (VSM) control emulates the dynamics and control of SGs and can provide inertia. Matching control structurally matches the differential equations of an inverter to those of an SG. However, virtual oscillator control (VOC) mimics the synchronizing behavior of Liénard-type oscillators and can globally synchronize an IBR-based power system. Furthermore, dispatchable VOC (dVOC) has been proposed to address the issue that the nominal power injection of VOC cannot be specified \cite{tayyebi2020frequency}. The structures of the most widely used droop control and VSM control are shown in Fig. \ref{GFM} \cite{pan2019transient}.
\begin{figure*}[htb]
  \centering
  \includegraphics[width=13.6cm]{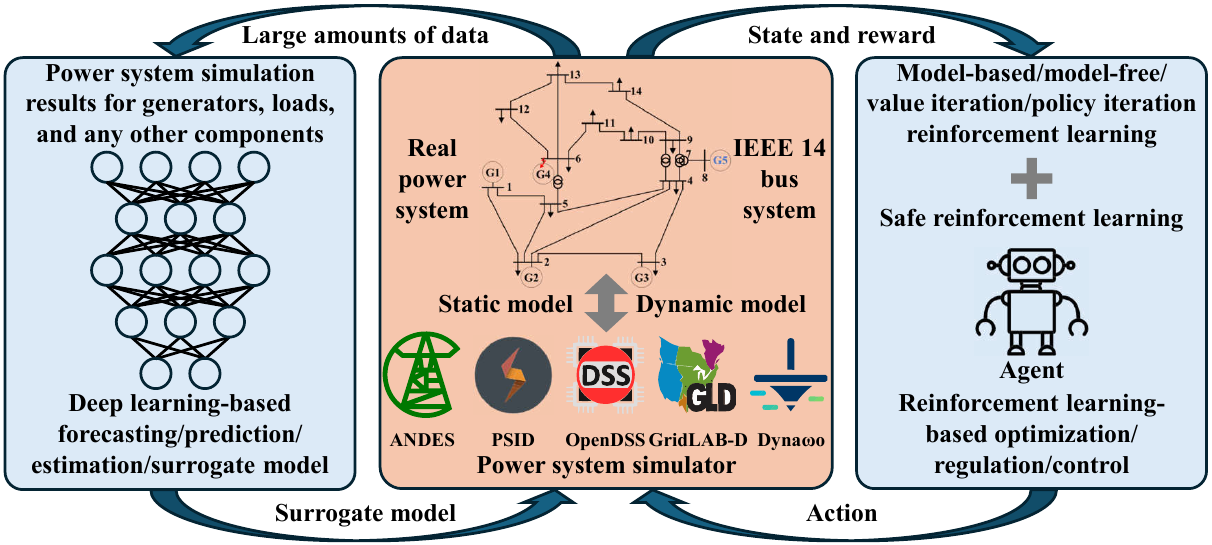}
  \caption{Framework of interaction between DL, RL, and dynamic simulators.}
  \label{Framework}
\end{figure*}
\subsection{Machine Learning Application}
Generally, the applications of machine learning in power systems can be divided into two categories: one focuses on forecasting, prediction, estimation, and surrogate modeling, primarily using DL; the other involves optimization, regulation, and control based on RL \cite{ibrahim2020machine, chen2022reinforcement}. DL methods typically require a large amount of data to accurately fit the dynamic trajectory of power systems following perturbations. However, since power systems operate with significant safety margins, dynamic data post-fault is difficult to obtain, and most existing research relies on data generated by power system simulators. To generate data, random or distribution-based Monte Carlo sampling of power system variables is necessary to cover a wide range of possible operating scenarios. These power system variables include SG and IBR active power outputs, SG and IBR bus voltage magnitudes, active and reactive power demands of loads, IBR control modes and parameters, as well as fault or perturbation locations and types, etc. RL are traditionally categorized into four types: model-based, model-free, value iteration, and policy iteration RL \cite{chen2022reinforcement}. When considering constraints, safe RL has been proposed to maximize rewards while ensuring constraint satisfaction through a series of specially designed techniques \cite{su2024review}. DL depends on dynamic simulators to perform TDSs for various scenarios and perturbations, generating large amounts of simulation data to train machine learning models. In contrast, RL requires extensive interaction with the environment, performing actions, receiving rewards and states, and iteratively training the model. The environment here can be a real-world system, a dynamic simulator, or a DL-based surrogate model. The interaction between DL, RL, and simulators is shown in Fig. \ref{Framework}.

Therefore, when applied to machine learning tasks, dynamic simulators must be flexible enough to allow users to easily modify all parameters through code, perform simulations, read results, and operate quickly. Typically, the time required to generate data is closely related to several factors including computer performance, the scale of the system, simulation duration, simulation step size, DAEs solver algorithm, programming language, complexity of the dynamic models, and the type of TDSs. For instance, QSP-based TDSs are significantly faster than EMT-based TDSs, and multi-GPU parallel processing is typically faster than single-GPU processing.

\section{Power System Dynamic Simulator}
\begin{table*}[!htb]
\renewcommand{\arraystretch}{1.1} %
\caption{Comparison of Dynamic Simulators: Base Functions and Models}
\label{Table_Comparison_1}
\centering
\begin{adjustbox}{center, max width=\textwidth}
\begin{tabular}{>{\centering\arraybackslash}m{2.3cm} >{\centering\arraybackslash}m{1.3cm} >{\centering\arraybackslash}m{0.5cm} >{\centering\arraybackslash}m{0.5cm} >{\centering\arraybackslash}m{0.7cm} >{\centering\arraybackslash}m{2.9cm} >{\centering\arraybackslash}m{5.5cm} >{\centering\arraybackslash}m{0.9cm} }
    \toprule %
    Simulator & Language & UM & PF & SSSA & TDS & RES model & Parallel\\
    \midrule %

    ANDES \cite{cui2020hybrid} & Python & \ding{55} & $\checkmark$ & $\checkmark$ & QSP (positive sequence) & REGCA1, REGCP1, REGCV1, REGCV2,  GFM-Droop, GFM-VSM, GFM-dVOC & $\checkmark$ \\

    PSID.jl \cite{lara2023powersimulationsdynamics} & Julia & \ding{55} & $\checkmark$ & $\checkmark$ & QSP (positive sequence)/ EMT (dq0-model) & REGCA1, GFM-Droop, GFM-VSM, GFM-dVOC & /\\
    
    Dyna$\omega$o \cite{guironnet2018towards} & C++/ Modelica & \ding{55} & $\checkmark$ & $\checkmark$ & QSP and quasi-EMT & WECC PV, WTG4A, WTG4B, GFM-Matching, GFM-Droop, GFM-dVOC & /\\

    OpenDSS \cite{dugan2011open} & Delphi & $\checkmark$ & $\checkmark$ & \ding{55} & QSP (positive sequence) & GFM-Droop, GFM-VSM, PV models, storage models, customized models (.dll) & / \\ 

    GridLAB-D \cite{chassin2014gridlab} & C/C++  & $\checkmark$ & $\checkmark$ & \ding{55} & QSP (positive sequence) & Basic PV, Battery, Wind, GFM-Droop (limited) & $\checkmark$ \\

    \bottomrule %
\end{tabular}
\end{adjustbox}
\begin{tablenotes}
    \item $\checkmark$, \ding{55} and / represent that the simulator has, does not have, and does not mention this function, respectively.
\end{tablenotes}
\end{table*}

\begin{table*}[!htb]
\renewcommand{\arraystretch}{1.1} %
\caption{Comparison of Dynamic Simulators: Machine Learning Applications}
\label{Table_Comparison_2}
\centering
\begin{adjustbox}{center, max width=\textwidth}
\begin{tabular}{>{\centering\arraybackslash}m{1.5cm} >{\centering\arraybackslash}m{4.2cm} >{\centering\arraybackslash}m{5.2cm} >{\centering\arraybackslash}m{2.6cm} >{\centering\arraybackslash}m{2.4cm}}
    \toprule %
    Simulator & Manual & Interaction with external code & Supported external data format & Simulation time (1/1000 samples) \\
    \midrule %

    ANDES & Detailed descriptions of all functions, models, and parameters & Easily implement Python-based parameter modification and result retrieval & MATPOWER, PSS/E, JSON & 1.27s/224s \\ 

    PSID.jl & Detailed descriptions of all functions, models, and parameters & Easily implement Julia-based parameter modification and result retrieval & MATPOWER, PSS/E, JSON & 0.05s/31.1s\\
    
    Dyna$\omega$o & Brief overview without descriptions of models or parameters & Executed via terminal command line, with limited code-based interaction & Unknown & 0.34s/374s\\ 


    OpenDSS & Comprehensive documentation on functions and models & Supports interaction with external scripts in Python and MATLAB for parameter modification and data retrieval & CSV, COMTRADE & / \\ 
    
    GridLAB-D & Detailed documentation with examples for functions and models & Supports integration with Python and C/C++ for extensive customization and external module development & CSV, XML, JSON  & / \\
    \bottomrule %
\end{tabular}
\end{adjustbox}
\begin{tablenotes}
    \item The simulation times for OpenDSS and GridLAB-D are not applicable (indicated as "/") because they are primarily designed for distribution networks.
\end{tablenotes}
\end{table*}
\subsection{ANDES}
ANDES is an open-source Python library for power system modeling, computation, and analysis, and it supports PF, TDS (QSP), and SSSA for transmission systems. ANDES implements a hybrid symbolic-numeric framework for rapid prototyping of DAE-based models, including the full second-generation renewable models: 
REGCA1 (Renewable energy generator model type A),
REGCP1 (Renewable energy generator model (REGC\_A) with PLL),
REGCV1 (Voltage-controlled voltage-source converter (VSC) with double-loop PI control and VSM control),
REGCV2 (Voltage-controlled VSC with lag transfer functions-based inner-loop current PI controllers and VSM control),
REGF1 (GFM-droop), REGF2 (GFM-VSM), and REGF3 (GFM-dVOC) \cite{cui2020hybrid}.


\subsection{PowerSimulationsDynamics.jl}
PowerSimulationsDynamics.jl (PSID.jl) is a Julia package designed for dynamic modeling of power systems with low inertia renewable energy sources (RESs) and features an extensive library of SG, inverter, and load models. It aims to provide a flexible modeling framework to accommodate various device models based on specific needs, streamline the construction of large-scale DAEs to enhance efficiency and reduce redundancy, and leverage Julia’s computational capabilities to boost performance in large-scale simulations. The supported TDS includes positive sequence QSP and dq0-model EMT \cite{lara2023powersimulationsdynamics}. The supported inverters include REGCA1, GFM-Droop, GFM-VSM, and GFM-dVOC \cite{henriquez2024small}.

\subsection{\texorpdfstring{Dyna$\omega$o}{Dynawo}}
Dyna$\omega$o has five submodules: DynaFlow for steady-state calculations, DySym for short-circuit calculations, DynaWaltz for long-term stability simulations, DynaSwing for short-term stability studies, and DynaWave for stability studies and system design with a high penetration of power electronics-based components (quasi-EMT). The supported inverters include WECC PV, WTG4A, WTG4B, GridFormingConverterMatchingControl (GFM-Matching), GridFormingConverterDroopControl (GFM-Droop) and GridFormingConverterDispatchableVirtualOscillatorControl (GFM-dVOC) \cite{guironnet2018towards}.


\subsection{OpenDSS}
OpenDSS is a powerful open-source tool for dynamic simulations in unbalanced distribution systems, widely used for high-penetration DERs, restoration actions, and GFM control. It supports three dynamic modes: (1) Default mode using built-in generator models (e.g., model=1 for second-order SGs or model=7 for IBRs), (2) DynamicExp function (available in OpenDSS v9) to define custom DAEs for advanced controls, and (3) User-written DLL files for flexible  modeling SGs and GFMs. These DLL files can be created in either the Delphi programming language or Free Pascal \cite{abiola2015getting}, allowing detailed modeling and control of internal structures and dynamic state variables. OpenDSS also includes a COM interface, enhancing integration with other software platforms. Configurations are specified in OpenDSS scripts with support for CSV and COMTRADE formats, and OpenDSS’s API compatibility with Python and MATLAB enables further customization \cite{dugan2011open}.

\subsection{GridLAB-D}
GridLAB-D is an open-source tool tailored for dynamic simulations in distribution systems, supporting TDS of  QSP models. It enables analysis of time-dependent behaviors such as load variability, demand response, and distributed energy resource integration, with basic models for PV, battery, wind systems, and GFM  with droop control for stability studies. GridLAB-D utilizes the GLM (GridLAB-D Model) format for model configurations and supports XML for data exchange, ensuring compatibility with external tools. Additionally, its Python and C/C++ API integration allows for extended dynamic simulation and custom module development \cite{chassin2014gridlab}.

We conducted a comprehensive comparison of these simulators based on two main aspects: (1) Base functions and model, including programming language, support for unbalanced modeling (UM), PF, SSSA, type of TDS, RES models, and parallel computing capabilities, as shown in Table \ref{Table_Comparison_1}; (2) Machine learning applications, focusing on whether they provide detailed manuals, allow flexible code-based parameter modification and result retrieval, supported external data formats, and their simulation time (IEEE 14-bus system; step size 0.01s; simulation duration 10s), as shown in Table \ref{Table_Comparison_2}. It should be noted that, due to Python's slower speed compared to Julia and C++, ANDES has the longest simulation time for a single run. However, with multi-CPU parallelization, the simulation time for 1,000 samples remains competitive.

\section{Simulation Comparison and Case Studies}
\begin{figure*}[htb]
  \centering
  \includegraphics[width=9cm]{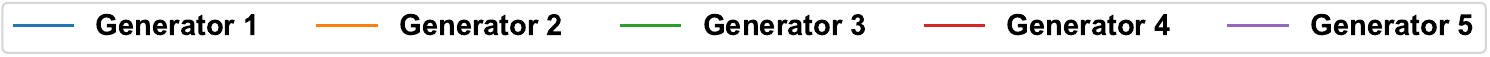}
  \includegraphics[width=9cm]{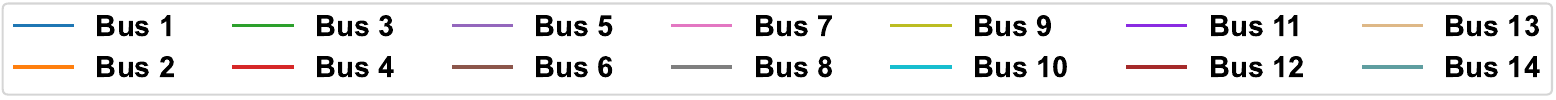}\\
  \includegraphics[width=4.45cm]{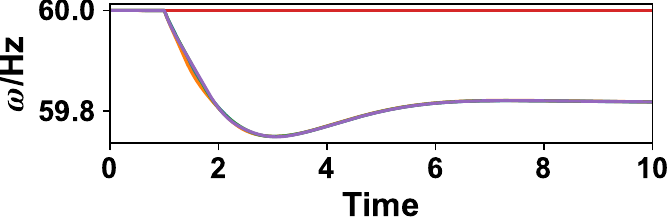}
  \includegraphics[width=4.45cm]{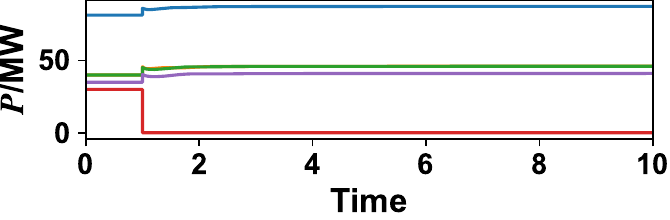}
  \includegraphics[width=4.45cm]{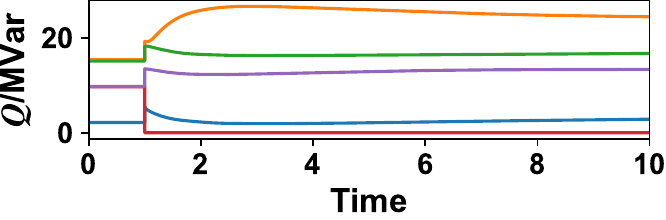}
  \includegraphics[width=4.45cm]{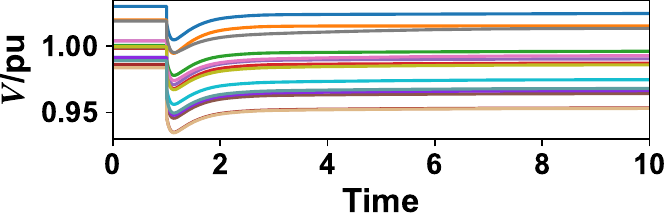}
  
  \includegraphics[width=4.45cm]{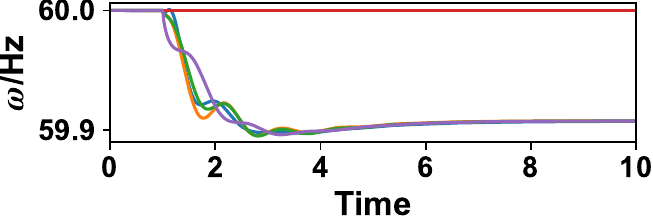}
  \includegraphics[width=4.45cm]{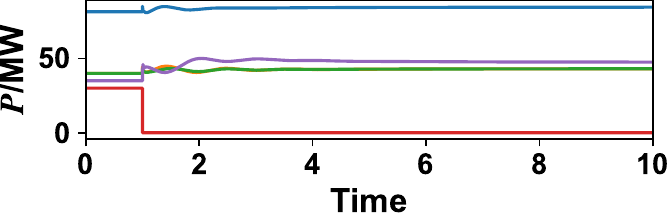}
  \includegraphics[width=4.45cm]{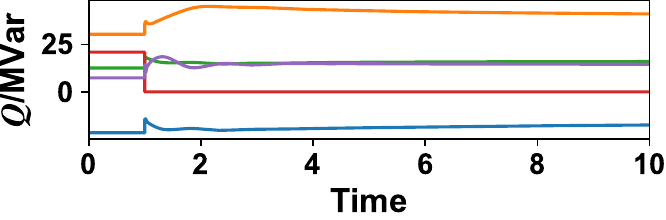}
  \includegraphics[width=4.45cm]{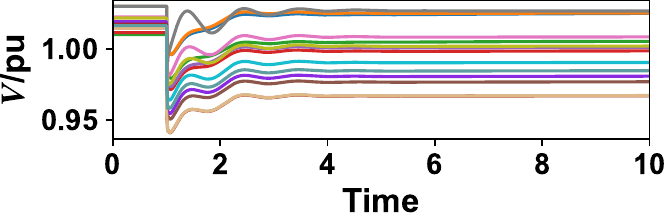}
  
  \includegraphics[width=4.45cm]{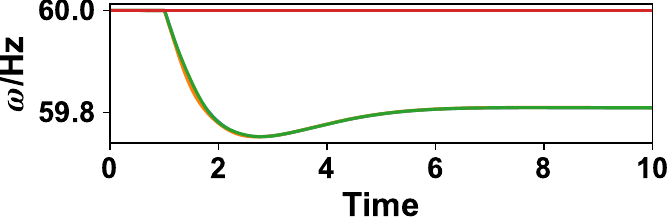}
  \includegraphics[width=4.45cm]{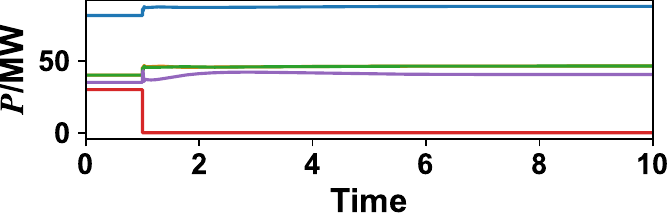}
  \includegraphics[width=4.45cm]{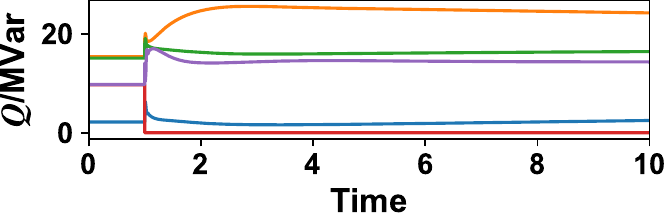}
  \includegraphics[width=4.45cm]{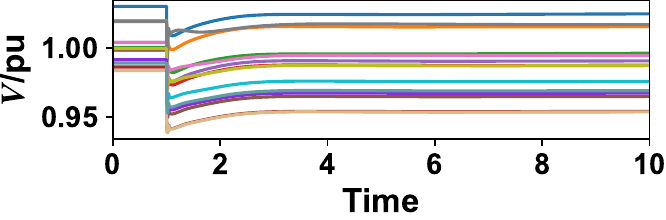}
  \caption{ANDES simulation results: from top to bottom, the three rows correspond to G5 configured as SG, GFM-Droop, and GFM-VSM, respectively; From left to right are the speed $\omega$, active power generation $P$, and reactive power generation $Q$ of the SG or GFM, followed by the voltage $V$ of all buses.}
  \label{ANDES}
\end{figure*}

\begin{figure*}[htb]
  \centering
  \includegraphics[width=4.45cm]{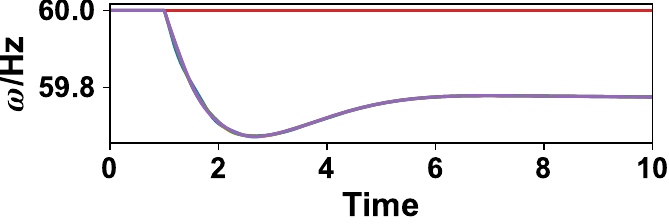}
  \includegraphics[width=4.45cm]{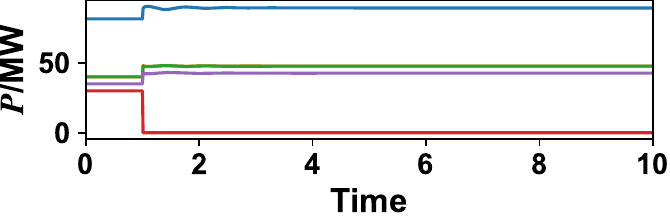}
  \includegraphics[width=4.45cm]{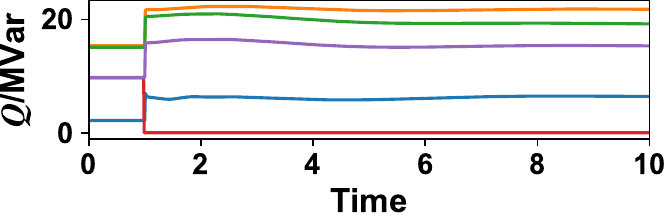}
  \includegraphics[width=4.45cm]{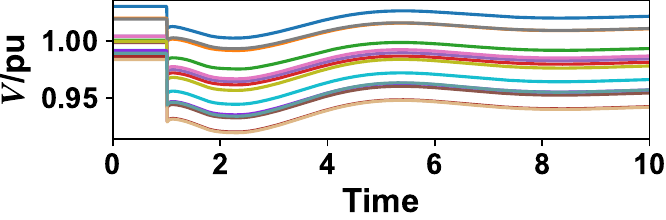}
  
  \includegraphics[width=4.45cm]{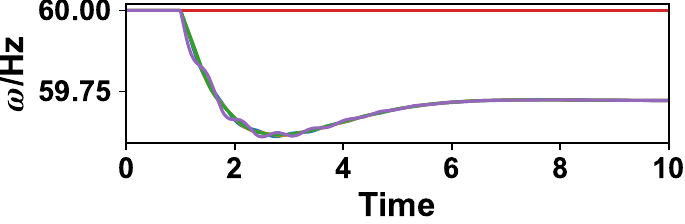}
  \includegraphics[width=4.45cm]{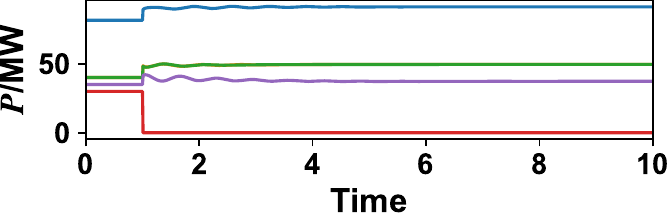}
  \includegraphics[width=4.45cm]{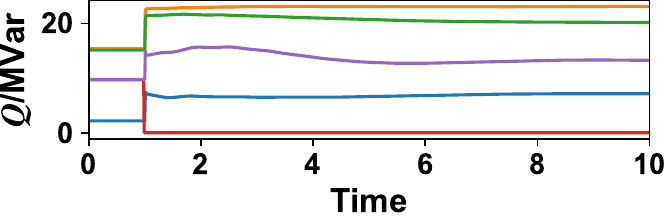}
  \includegraphics[width=4.45cm]{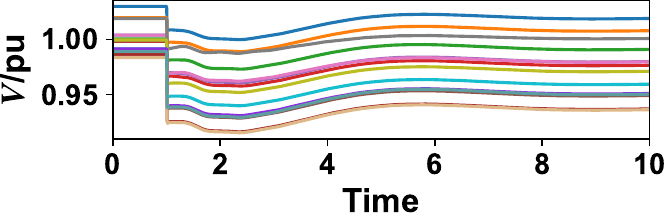}
  
  \includegraphics[width=4.45cm]{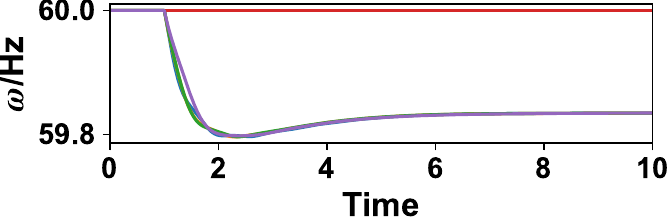}
  \includegraphics[width=4.45cm]{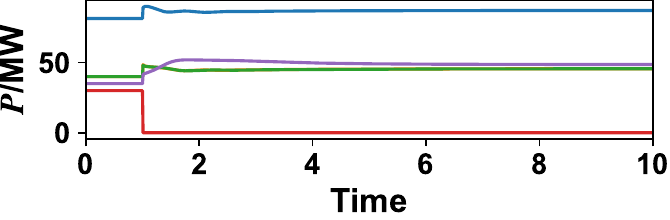}
  \includegraphics[width=4.45cm]{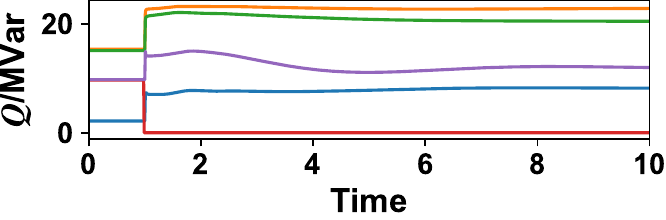}
  \includegraphics[width=4.45cm]{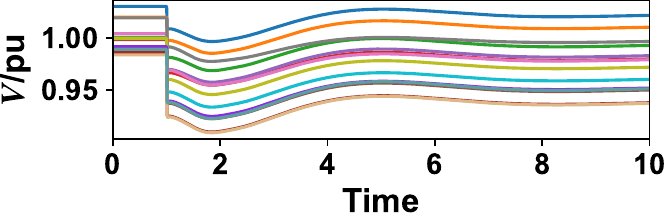}
  \caption{PSID.jl simulation results: from top to bottom, the three rows correspond to G5 configured as SG, GFM-Droop, and GFM-VSM, respectively; From left to right are the speed $\omega$, active power generation $P$, and reactive power generation $Q$ of the SG or GFM, followed by the voltage $V$ of all buses.}
  \label{PSID}
\end{figure*}

Since ANDES and PSID.jl provide detailed manuals and programming guidance, TDSs were performed on these two simulators using a modified IEEE 14-bus system, where generators 1-4 are SGs and generator 5 is set as SG, GFM-Droop, and GFM-VSM sequentially. The fault is configured as a generation trip of generator 4, beginning at 1s, with a total simulation duration of 10s. The modified 14-bus system structure, ANDES simulation results, and PSID.jl simulation results are shown in Fig. \ref{Framework}, \ref{ANDES}, and \ref{PSID}, respectively. The simulation results of the two simulators are very similar, with differences arising from variations in dynamic models and control parameters.

\section{Conclusion}
This paper presents an in-depth comparison of the latest open-source dynamic simulators that support GFM inverters, with a focus on their suitability for machine learning applications.
This survey offers machine learning professionals practical guidance on power system data generation and interaction with dynamic simulators.

\section*{Acknowledgments}
The views expressed in the article do not necessarily represent the views of the DOE or the U.S. Government. The U.S. Government retains, and the publisher, by accepting the article for publication, acknowledges that the U.S. Government retains a nonexclusive, paid-up, irrevocable, worldwide license to publish or reproduce the published form of this work or allow others to do so for U.S. Government purposes.

\bibliographystyle{IEEEtran}
\bibliography{ref.bib}

\end{document}